\documentclass[preprint,12pt]{elsarticle}
\newcommand{\R}{{\mathbb{R}}}

\newcommand{\N}{{\mathbb{N}}}
\usepackage{amssymb}

\journal{Annals of Physics}

\begin{document}
%\sloppy
%\draft
\begin{frontmatter}

\title{ A phenomenon of  
splitting  resonant-tunneling one-point interactions 
 }
\author{A.V. Zolotaryuk}
%\email{azolo@bitp.kiev.ua}
\address
{Bogolyubov Institute for Theoretical Physics, National Academy of
Sciences of Ukraine, Kyiv 03143, Ukraine}

\date{\today}

\begin{abstract}

The so-called $\delta'$-interaction as a particular example in 
 Kurasov's  distribution theory developed on the space of  discontinuous 
(at the point of singularity) test functions,  is identified with the diagonal
transmission matrix,  {\em continuously} depending on the strength of this interaction.
On the other hand, in several recent publications, the $\delta'$-potential 
has been shown to be transparent at some {\em discrete} values of the strength 
constant and opaque beyond these values. 
This discrepancy is resolved here on the simple physical example, namely
the  heterostructure consisting of two extremely thin layers separated by infinitesimal
distance.   In the three-scale squeezing limit as
the thickness of the layers and the distance between them simultaneously
tend to zero,  a whole variety of single-point interactions is realized.
The key point is the generalization of the $\delta'$-interaction to 
the family for which 
 the resonance sets appear in the form of a countable number of continuous 
two-dimensional curves. In this way, the connection between Kurasov's $\delta'$-interaction
and the resonant-tunneling point interactions is derived and the splitting of the resonance sets  for tunneling plays a crucial role. 
  \end{abstract}

%\pacs{03.65.-w, 03.65.Nk, 73.40.Gk}
%\maketitle

\end{frontmatter}

%===========================================Introduction================================
\section{Introduction}
%===========================================Introduction================================

Starting with the pioneering work by Berezin and Faddeev \cite{bf}, various 
exactly solvable models 
described by the Schr\"{o}dinger operators with singular zero-range potentials have
 been studied within the theory of selfadjoint extensions of symmetric operators.
These models are specified by the  potentials  defined on the sets 
consisting of isolated points and therefore in the literature they are usually referred to as 
``point interactions'' (see monographs \cite{do,a-h,ak} for
details and references). According to this theory, all the selfajoint extensions of 
the kinetic energy operator form a four-parameter family \cite{adk,afk}, so that there are 
different ways to define the limit Schr\"{o}dinger operator being appropriate for a given physical system. A whole body of literature (see, e.g., 
\cite{k,cnp,cnt,cft,tfc,an,gnn,ggn}, a few
to mention), including the very recent studies 
\cite{l1,bn,gggm,l2,kp,gmmn,knt1,knt2} 
with references therein, has been published  
where  the  one-dimensional  Schr\"{o}dinger operators were defined via  
distributions 
and corresponding two-sided boundary conditions at the points of singularity.  
The advantage of this ``point'' approach is the possibility to get 
 the resolvents of these operators in an explicit form, to find their spectra, to compute scattering coefficients, etc.  

On the other hand, the distributional part of Schr\"{o}dinger operators can be treated as the limit of regularized potentials.
Within this approach different asymptotic methods are used for realizing limit point interactions. Particularly, in dimension one, 
the regularized stationary Schr\"{o}dinger equation  
%----------------------------------------1---------------------------
\begin{equation}
-\psi''(x) +V_\varepsilon(x)\psi(x)= E \psi(x),
\label{1}
\end{equation}
%----------------------------------------1---------------------------
where  the prime stands for the derivative with respect to the spatial coordinate $x$ and 
$\psi(x)$ is the wavefunction of a particle with energy $E$, has been used. 
The potential $V_\varepsilon(x)$  is supposed to depend on the squeezing parameter 
$\varepsilon > 0$, so that
in the limit as $\varepsilon \to 0$, the function $V_\varepsilon(x)$ is confined to
one point. Using the asymptotic approach, most of papers 
\cite{s,c-g,zci,tn,gm,zpla10,gh,gh1} have been devoted to 
studying the  interactions of the point dipole type which are 
realized in the limit $V_\varepsilon(x) \to \gamma \delta'(x)$ in the sense of distributions
($\gamma \in \R$ is a coupling constant). 
In addition, the Schr\"{o}dinger operators with $(a\delta' + b\delta)$-like potentials
have been investigated in a series of publications \cite{gnn,ggn,zz11,g1,g2}.

In the important work \cite{k}, Kurasov has developed the distribution theory based
on the space of discontinuous at the point of singularity (say, at $x=0$) test functions.
Within this theory, it is possible to define rigorously, as a particular example,
a point interaction referred in the following to as Kurasov's $\delta'$-interaction,
which is determined by the one-parameter transmission matrix 
%---------------------------------------2-----------------------------------------
\begin{eqnarray}
 \left( \begin{array}{cc} \psi(+0) \\
 \psi'(+0) \end {array} \right) = \Lambda \left( \begin{array}{cc} \psi(-0) \\
 \psi'(-0) \end {array} \right), ~~ \Lambda =  \left(
\begin{array}{cc} \theta  ~ ~~~ 0 ~~\\
 0~~ ~~ \theta^{-1} \end {array} \right),~~\theta = {2+ \gamma \over 2 - \gamma }
\label{2}
\end{eqnarray}
%-----------------------------------------2-----------------------------------------------
where the parameter
 $\gamma \in  \R\setminus \{ \pm 2 \}$ serves as a coupling constant of this interaction.
This transmission matrix has widely been used by many authors 
(see, e.g., \cite{gnn,l1,bn,gggm,l2,kp, gmmn}).
On the other hand, 
beginning from the  paper \cite{gm}, for a whole class of conventional approximations
of the $\delta'$-potential, Golovaty and coworkers 
 have rigorously established the existence of discrete resonance sets in the $\gamma$-space
on which the tunneling through this point barrier appears to be non-zero, whereas beyond 
these sets the system is fully opaque. Moreover, on the resonance sets they have 
 developed the procedure how to compute the transmission matrix for this tunneling. 
This type of point interactions may be referred to as 
{\em resonant-tunneling} $\delta'$-potentials. Note that 
the only common feature of Kurasov's  $\delta'$-interaction and the resonant-tunneling
$\delta'$-potential is that the
transmission matrices  of both these interactions are of the diagonal form. 
In this regard, it is important to develop an approach within which 
both these types of interactions could  somehow be connected.
 Therefore 
 the goal of the present paper is to realize both these types of  interactions
within a unique description starting from the {\it same} profile of the potential 
$V_\varepsilon(x)$ in Eq.\,(\ref{1}). 

It is fascinating that the connection between Kurasov's  $\delta'$-interaction
and the family of resonant-tunneling $\delta'$-potentials can be described 
on the basis of the most simple physical system.  
We show that Kurasov's $\delta'$-potential emerges from the realistic heterostructure
 consisting of two  thin parallel plane layers separated by some distance
in the limit as  both the thickness of layers and
the distance between them simultaneously tend to zero in a certain way.
In other squeezing limits,
the limit one-point interactions are proved to depend crucially on the relative coming up to
zero of the thickness and the distance. As a result, a whole variety of single-point interactions
occurs in this limit depending on the way of convergence. Surprisingly, within this
approach, it is possible to realize both the Kurasov $\delta'$-interaction
and the family of $\gamma \delta'$-potentials with countable sets in the $\gamma$-space 
at which a non-zero resonant tunneling takes place. The key point is that we have 
to extend the family of $\delta'$-potentials to wider class of interactions for which
the resonance sets become curves instead of points.
Another surprising point is that
the  $\delta$-potential discovered by \v{S}eba in \cite{s} can also be realized
under a certain way of squeezing.

In general, one can consider the structure consisting of arbitrary $N$ separated layers.
Then the potential in Eq.\,(\ref{1}) can be expressed as a piecewise constant function
depending on  barrier heights
or  well depths $h_j \in \R$, widths $l_j$, $j= \overline{1,  N}$, and the 
distances between the layers $r_j$, $j=\overline {1, N-1}$,
 such that $|h_j| \to  \infty$ and $l_j,\,r_j \to 0$ as $\varepsilon \to 0$.
Using the power-connecting parametrization $h_j = a_j \varepsilon^{-\mu_j}$,
$r_j =c_j \varepsilon^{\tau_j}$ ($c_j >0$) with positive powers $\mu_1, \ldots , \mu_N$
and $\tau_1, \ldots , \tau_{N-1}$, where $a_j \in \R$ may be called
characteristic {\it intensities} of the layers, the potential $V_\varepsilon(x)$ can be represented in the form of the function 
$V_\varepsilon(l_1, \ldots , l_N; a_1, \ldots , a_N; \mu_1, \ldots , 
\mu_N; \tau_1, \ldots , \tau_{N-1}; x)$. The problem to be solved is the finding 
of the conditions on the parameters
$a_1, \ldots , a_N$ and $\mu_1, \ldots , \mu_N; \tau_1, \ldots , \tau_{N-1}$
at which all the possible families of point interactions can be realized in the limit as $\varepsilon \to 0$.

The most simple situation appears if $\mu_1= \ldots = \mu_N \equiv \mu$ and
$ \tau_1 = \ldots = \tau_{N-1} \equiv \tau$. The two cases of a 
double- and a triple-layer structures  
($N=2,3$) have been analyzed in detail in the recent work \cite{z17}. Here the 
parameter $\mu$ controls the {\it same} rate of shrinking the layers, whereas
the parameter $\tau$ describes the rate of decreasing the distance between the layers.
Various families of single point interactions have been realized on different
 two-dimensional $\{\mu, \tau \}$-sets. In the present work we assume the {\it different}
shrinking of two layers, i.e., $\mu_1 \equiv \mu$ and $\mu_2 \equiv \nu$, so that the
power-like connection occurs here between the three parameters: $\mu$, $\nu$ and $\tau$. 
On the other hand, to keep things simple, we restrict ourselves only 
to a double-layer structure, but enlarge the number of squeezing parameters from two
to three. In this (three-dimensional) space, we will find the open sets where 
Kurasov's $\delta'$-interaction as well as \v{S}eba's $\delta$-potential \cite{s} 
are defined. Within these sets, under approaching their limiting sets, 
the {\em splitting} of these interactions into  countable families of one-point interactions 
is shown to occur 
and the description of this phenomenon is the key point of the present paper.

The paper is organized as follows. In Section 2 we present the piecewise 
constant potential for the double-layer structure and the transmission 
matrix for this system. The conditions for the resonant tunneling through the
double-layer system in the limit as the layer thickness squeezes to one point
are derived in Section 3 in a general form. In Section 4 we introduce a two-scale
power-connecting parametrization of the layer parameters and describe the splitting
of three types of point interactions. The additional parametrization of the distance between 
the layers is present in Section 5 for the visualization of the splitting effect and
\v{S}eba's transition. 
The paper is concluded by Section 6 in which we summarize the results 
with the discussion of possible extensions.

%===========================================2====================================
\section{Finite-range potential and its transmission matrix}
%===========================================2====================================

Consider the system consisting of two separated layers described by the 
piecewise constant potential 
%----------------------------------------3--------------------------
\begin{equation}
\bar{V}(h_1,h_2,l_1,l_2,r;x)= \left\{ \begin{array}{ll}
 h_1 &   \mbox{for}~~ 0< x < l_1 , \\
 h_2     & \mbox{for}~~ l_1+r < x < l_1+ r+l_2 ,\\
0 & \mbox{for}~ -\infty < x <0, ~l_1< x< l_1+r,\\
  & ~~~~~~~l_1+r+l_2 < x < \infty ,
\end{array} \right. 
\label{3}
\end{equation} 
%--------------------------------------3-----------------------------
where $h_j \in \R$ ($h_j > 0$, barrier; $h_j < 0$, well), $l_j >0$ (layer thickness), $r >0$ (distance between layers), $j=1,2$.  
  The transmission matrix $\bar{\Lambda}$ 
for Eq.\,(\ref{1}) with this potential is defined by the relations
%---------------------------------------4--------------------------------
\begin{eqnarray}
\left( \begin{array}{cc} \psi(x_2)  \\
\psi'(x_2) \end{array} \right) 
 = \bar{\Lambda} \left(
\begin{array}{cc} \psi(x_1)   \\
\psi'(x_1)   \end{array} \right), ~~~ \bar{\Lambda}= 
 \left( \begin{array}{cc} \bar{\lambda}_{11}~~ \bar{\lambda}_{12} \\
\bar{\lambda}_{21} ~~\bar{\lambda}_{22} \end{array} \right) .
\label{4}
\end{eqnarray}
%-----------------------------------------4-------------------------------
It connects the boundary conditions of the wave function
$\psi(x)$ and its derivative $\psi'(x)$ at $x=x_1= 0 $ 
and $x=x_2 =  l_1+r+l_2  $. 
The notations with the overhead bars have been introduced for the finite-range
quantities. Explicitly, the elements of the $\bar{\Lambda}$-matrix 
that corresponds to the potential (\ref{3}) are given by
%------------------------------------------5-8--------------------------------
\begin{eqnarray}
\bar{\lambda}_{11} & = & \left[ \cos(k_1l_1) \cos(k_2l_2) 
-(k_1/k_2)\sin(k_1l_1) \sin(k_2l_2) \right] \cos(kr)
\nonumber \\
& -& \left[ (k_1/k) \sin(k_1l_1) \cos(k_2l_2) +   
(k/k_2) \cos(k_1l_1) \sin(k_2l_2) \right] \sin(kr), 
\label{5}\\
\bar{\lambda}_{12} & = & \left[(1/k_1) \sin(k_1l_1) \cos(k_2l_2)
+ (1/k_2)\cos(k_1l_1) \sin(k_2l_2) \right] \cos(kr)
\nonumber \\
& +& \left[ (1/k) \cos(k_1l_1) \cos(k_2l_2)
  -  (k/k_1k_2) \sin(k_1l_1) \sin(k_2l_2) \right] \sin(kr), ~~~~
\label{6}\\
\bar{\lambda}_{21} & = & - \left[ k_1 \sin(k_1l_1)\cos(k_2l_2)
+ k_2 \cos(k_1l_1)\sin(k_2l_2) \right] \cos(kr)
\nonumber \\
& -& \left[ k \cos(k_1l_1) \cos(k_2l_2)  
-  (k_1k_2/k) \sin(k_1l_1) \sin(k_2l_2) \right] \sin(kr), 
\label{7}\\
\bar{\lambda}_{22} & = & \left[ \cos(k_1l_1) \cos(k_2l_2)
-(k_2/k_1)\sin(k_1l_1) \sin(k_2l_2) \right] \cos(kr)
\nonumber \\
& -& \left[ (k/k_1) \sin(k_1l_1)\cos(k_2l_2)
 +   (k_2/k) \cos(k_1l_1) \sin(k_2l_2) \right] \sin(kr), 
\label{8}
\end{eqnarray}
%------------------------------------------5-8-------------------------------------
where 
\begin{equation}
 k_j := \sqrt{E -h_j}\,,~~ j=1,2,~~k := \sqrt{E}\,.
\label{9}
\end{equation}

%================================================3================================
\section{Squeezing limit: Resonance conditions}
%===================================================3===============================

 The squeezing limit of the system given by the potential (\ref{3}) means that 
$l_j, r \to 0$ but $|h_j| \to \infty$, $j=1,2$.  
Therefore if the matrix elements (\ref{5})\,-\,(\ref{8}) are 
finite in the squeezing limit, we adopt the following notations:
%------------------------------------------10-----------------------------------
\begin{equation}
  \bar{\Lambda} \to \Lambda =
\left( \begin{array}{cc} {\lambda}_{11}~~ {\lambda}_{12} \\
{\lambda}_{21} ~~{\lambda}_{22} \end{array} \right),~~~\bar{\lambda}_{ij} \to
{\lambda}_{ij},~~i,j =1,2, 
\label{10}
\end{equation}
%--------------------------------------------10------------------------------------
where the limit elements are denoted without overhead bars.
Next, having  accomplished the limit procedure,  we set $x_1 =-\,0$ and 
$\lim_{l_j, r \to 0}x_2 = +\,0$. 

%===================================================3.1===============================
\subsection{Two particular cases of point interactions}
%===================================================3.1================================

Consider some trivial cases of the convergence of the $\bar{\lambda}_{ij}$-elements 
given by Eqs.\,(\ref{5})\,-\,(\ref{8}) in the limit as
$l_1, l_2, \sin(kr) \to 0$. The first of these is a $\delta$-like profile of the 
layers. In this case, we have to assume in the squeezing limit that 
$h_jl_j =\alpha_j \in \R$, $j=1,2$. Then  
$k_j \to \sqrt{-\, h_j} \to \sqrt{-\,\alpha_j/ l_j}$ and, as a result, the limit transmission matrix 
becomes 
%------------------------------------------11-----------------------------------
\begin{equation}
  \Lambda =
\left( \begin{array}{lr} ~~~1 ~ ~~~~~~0 \\
\alpha_1 +\alpha_2  ~~1 \end{array} \right),
\label{11}
\end{equation}
%--------------------------------------------11-----------------------------------
which describes the $\delta$-interaction with the coupling constant that equals the algebraic sum of the layer intensities.

The second case, which also follows from  Eqs.\,(\ref{5})\,-\,(\ref{8}), 
concerns with a double-well structure ($h_j \le 0$). 
Here the transmission across 
the system occurs perfect ($\Lambda =\pm I$, $I$ is the identity matrix)
if $\sin(k_jl_j) \to 0$ or if $k_1 =k_2$ and $\cos(k_jl_j) \to 0$. As a result,
 we obtain the following two types of
conditions on the system parameters:
%----------------------------------------------12------------------------------------
\begin{equation}
\sqrt{-\, h_1}\, l_1 = m\pi, \sqrt{-\, h_2}\, l_2 =n\pi
~~\mbox{and}~~h_1=h_2,\sqrt{-\, h_1}\, l_1 = (n +1/2)\pi
\label{12}
\end{equation}
%--------------------------------------------------12----------------------------------
with $m,n =0,1, \ldots.$

\subsection{Three types of resonant-tunneling point interactions}

The arguments $k_jl_j$ of the trigonometric functions in Eqs.\,(\ref{5})\,-\,(\ref{8})
must be finite under the squeezing of the widths $l_1$ and $l_2$.
Therefore since $|h_j| \to \infty$, then $|k_j| \to \infty$, $j=1,2$. 
We assume that $k_j l_j \to A_j $, where $A_j$'s are required to
 be zero or finite 
non-zero (either real or imaginary) constants.
For the realization of resonant-tunneling (connected) point interactions,
the elements of the limit ${\Lambda}$-matrix must be finite in 
the squeezing limit.
As can be seen from the explicit representation (\ref{5})\,-\,(\ref{8}),
 the element $\bar{\lambda}_{21} $ appears to be  the most singular term
in this limit. Hence, we have to  assume the following limit:
%-----------------------------------------13-----------------------------------
\begin{equation}
\bar{\lambda}_{21} \to \alpha,
\label{13}
\end{equation}
%-----------------------------------------13-----------------------------------
where $\alpha \in \R$ is an arbitrary constant.
There are two ways of cancellation of divergences in the element 
$\bar{\lambda}_{21}$ as the layers are squeezed to zero and 
three types of connected point interactions can be realized as follows.

(i): One of the ways of cancellation of divergences in the $\bar{\lambda}_{21}$-element
[see Eq.\,(\ref{7})] that provides the limit (\ref{13}) is the asymptotic equation
%----------------------------------------------14----------------------------------------
\begin{equation}
{\tan(kr) \over k}= {l_1 \over A_1}\cot\!A_1 + {l_2 \over A_2}\cot\!A_2\,.
\label{14}
\end{equation}
%------------------------------------------------14---------------------------------------
It follows from this equation that  $\sin(kr) \to 0$ as $l_1, l_2 \to 0$ and therefore 
using this limit in Eq.\,(\ref{7}), we find that $\alpha =0$ in Eq.\,(\ref{13}). 
Using  next the resonance condition (\ref{14}), one can find the asymptotic representation
of the diagonal elements of the $\bar{\Lambda}$-matrix. Thus,
inserting this condition into the expressions (\ref{5}) and (\ref{8}) 
for $\bar{\lambda}_{11}$ and $\bar{\lambda}_{22}$, we obtain the following three 
asymptotic representations:
%----------------------------------------------------15---------------------------------
\begin{eqnarray}
\bar{\lambda}_{11} \,,\,  \bar{\lambda}_{22}^{-1} 
& \to & {\cos\!A_1 - (A_1/kl_1)\sin\!A_1 \tan(kr)\over \cos\!A_2} \nonumber \\
& & =  {\cos\!A_1 \over \cos\!A_2 - (A_2/kl_2)\sin\!A_2 \tan(kr)} 
= -\, {A_1 l_2 \sin\!A_1 \over A_2 l_1 \sin\!A_2}\, .
\label{15}
\end{eqnarray}
%--------------------------------------------------------15------------------------------

(ii):
The second way of cancellation of divergences in the $\bar{\lambda}_{21}$-element
is to assume  the equation 
%-------------------------------------------------16-------------------------------------
\begin{equation}
 {l_1 \over A_1}\cot\!A_1 + {l_2 \over A_2}\cot\!A_2 =0 .
\label{16}
\end{equation}
%--------------------------------------------------16------------------------------------
 In this case the limit (\ref{13}) reduces to 
%--------------------------------------------------17-----------------------------------
\begin{equation}
 {A_1 A_2 \over l_1l_2}
 \sin\!A_1 \sin\!A_2 {\sin(kr) \over k} \to \alpha \, .
\label{17}
\end{equation}
%------------------------------------------------------17----------------------------------
For a given $\alpha \in \R$, from this limit we find the second dependence of     
$\sin(kr)$ on the widths $l_1, l_2$: 
%----------------------------------------------------18---------------------------------
\begin{equation}
{\sin(kr) \over k}= {\alpha l_1 l_2 \over A_1 A_2  \sin\!A_1 \sin\!A_2 }\,.
\label{18}
\end{equation}
%----------------------------------------------------18------------------------------------
Similarly, using the resonance condition (\ref{16}) as well as the dependence (\ref{18}), we find 
 the limit diagonal elements of the ${\Lambda}$-matrix
in the second case of cancellation of divergences: 
%-------------------------------------------------19---------------------------------
\begin{equation}
\bar{\lambda}_{11} \, , \, \bar{\lambda}_{22 }^{-1}
 \to {\cos\!A_1 \over \cos\!A_2} 
= -\, {A_1 l_2 \sin\!A_1 \over A_2 l_1 \sin\!A_2}\, .
\label{19}
\end{equation}
%----------------------------------------------------19------------------------------

(iii): Finally, the third type of resonant-tunneling point interactions can be realized
on the same resonance set defined by Eq.\,(\ref{16}), however, for this type we assume
 that the limit $\sin(kr) \to 0$ proceeds faster (as a function of $l_1, l_2$)
than in the asymptotic representation (\ref{18}). The limit diagonal elements 
$\lambda_{11}$ and $\lambda_{22}$ are given in this case by the same formulae
(\ref{19}).

%==============================================4====================================

\section{A power-connecting representation of the  layer parameters 
}
%==============================================4===================================

One of the ways to study 
the convergence  $\bar{\Lambda} \to \Lambda$ as $l_1, l_2 \to 0$
is the  connection of $l_j$ and $h_j$, $j=1,2$, 
through the squeezing parameter $\varepsilon >0$
 using two  positive powers $\mu $ and $\nu$   as follows 
%-------------------------------------------------20--------------------------------
\begin{equation}
l_1=\varepsilon, ~~ l_2=\eta \varepsilon^{1-\mu +\nu}~(0 < \eta < \infty),
~~ h_1 = a_1 \varepsilon^{-\mu},~~h_2 = a_2 \varepsilon^{-\nu}.
\label{20}
\end{equation}
%-------------------------------------------------20-----------------------------------
Here the 
coefficients $a_j \in \R$, $j=1,2$, are characteristic quantities  of the system, 
so that they may be called  the intensities  of
 layers. Inserting the parametrization (\ref{20}) into the potential $\bar{V}$ 
[see Eq.\,(\ref{3})]
and the matrix $\bar{\Lambda}$, we replace the notations as $\bar{V}(x) \to V_\varepsilon(x)$
and $\bar{\Lambda} \to \Lambda_\varepsilon$.
Because of the limit $l_2 \to 0$, the inequality $1-\mu +\nu >0$ is necessary. 
We assume $\mu >1$ because for $\mu <1$ the transmission is trivially perfect
and the case $\mu =1$ reduces to the $\delta$-potential with the transmission 
matrix (\ref{11}). 

%===================================================4.1===========================
\subsection{Sets of the existence of the distribution $\delta'(x)$}
%====================================================4.1=========================

\begin{figure}
\centerline{\includegraphics[width=1.0\textwidth]{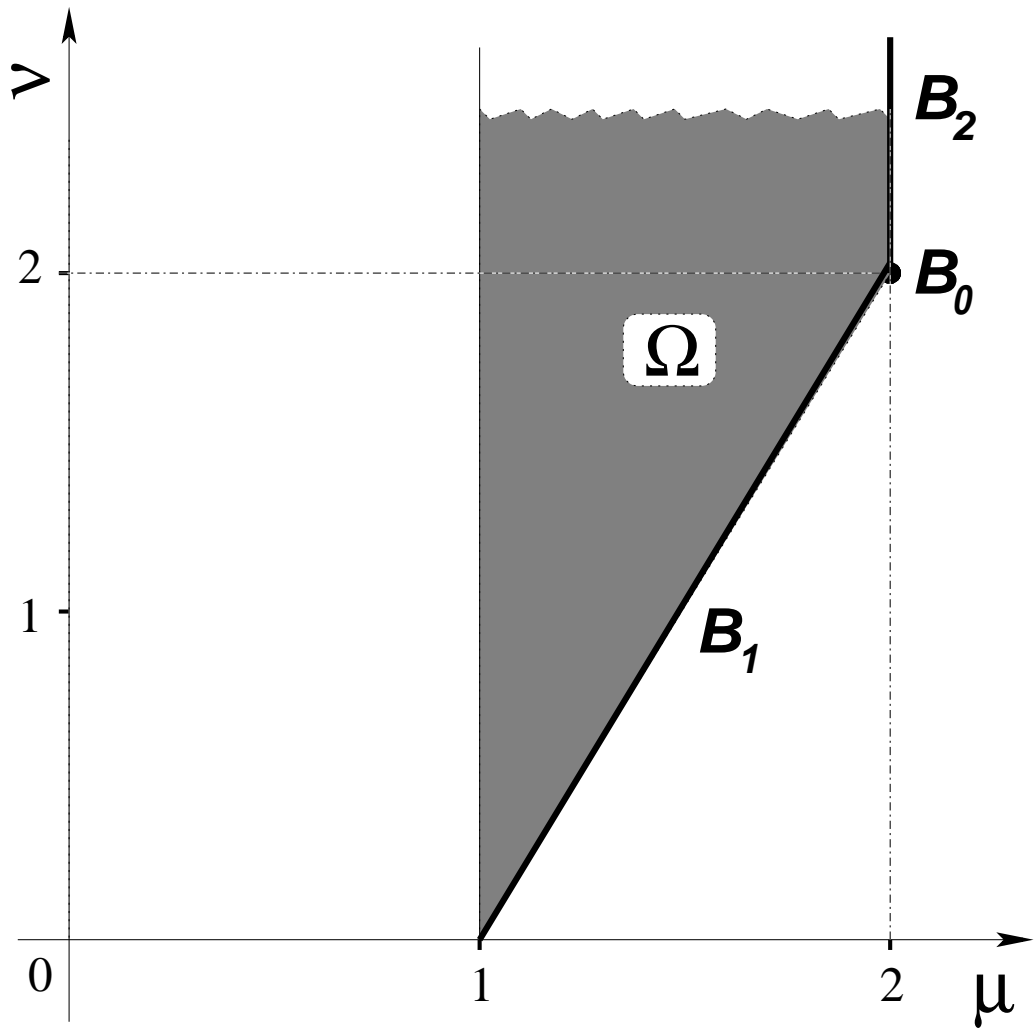}}
%\vspace{2pt}
\caption{
Schematics of the $\Omega$-set and its limiting (boundary) sets: 
point $ B_0$, lines $B_1$ and $B_2$, defined by Eqs.\,(\ref{21}) and (\ref{22}). 
}
\label{fig1}
\end{figure}

One can prove that the parametrized function $V_\varepsilon(x)$ converges 
 in the sense of distributions to the derivative delta potential $\gamma \delta'(x)$ with the coupling constant $\gamma$ given below. 
 This convergence  takes place on the sets  $ B_j$, $j=0,1,2$, 
 shown in Fig.\,\ref{fig1} in the case if $r=0$
and  $a_1 +\eta a_2 =0$ with arbitrary positive $\eta$. 
%---------------------------------------------fig1-------
%-------------------------------------------------fig1------------- 
and defined by
%-----------------------------------------------21--------------------------------------------
\begin{eqnarray}
\left. \begin{array}{ll} 
B_0  := \{ \mu = \nu =2  \},  \\
 B_1  := \{ 1 <\mu < 2, \, \nu =2(\mu -1) \},  \\
B_2  := \{ \mu =2, \, 2 < \nu < \infty \},  \end{array} \right.
\label{21}
\end{eqnarray}
%-----------------------------------------------21--------------------------------------------
which are the limiting sets of the open set 
%-----------------------------------------------22-------------------------------------------
\begin{equation}
\Omega  :=  \{ 1 <\mu < 2, \, 2(\mu -1 )< \nu < \infty  \}.
\label{22}
\end{equation}
%---------------------------------------------------22--------------------------------------
The coupling constant $\gamma$ of the $\delta'$-potential is the set function of $B_j$'s:
%----------------------------------------------------23---------------
\begin{equation}
\gamma = \gamma(B_j)= {a_1 \over 2} \left\{ \begin{array}{lll}
 1+\eta    & \mbox{for}~ B_0 ,\\
 \eta  & \mbox{for}~ B_1 ,\\
1  & \mbox{for}~ B_2.
\end{array} \right. 
\label{23}
\end{equation} 
%---------------------------------------------------23----------------

In the limit as $\varepsilon \to 0$, from Eqs.\,(\ref{9}) and (\ref{20}) we get
the asymptotic representation 
%--------------------------------------------24---------------------------------------
\begin{equation}
\left. \begin{array}{ll} k_1 \to \sqrt{-\,a_1} \,
\varepsilon^{-\mu /2}, ~~~ &  k_2 \to  \sqrt{-\,a_2} \,
\varepsilon^{-\nu /2}, \\
 A_1 =  \sqrt{-\, a_1} \, \varepsilon^{1-\mu/2 }, ~ ~~ & A_2 = \eta \sqrt{-\, a_2} 
\, \varepsilon^{1-\mu + \nu/2 }. \end{array} \right.
\label{24}
\end{equation}
%--------------------------------------------24--------------------------------------- 
On the sets (\ref{21}) and (\ref{22}), the asymptotic formulae 
(\ref{24}) provide the finiteness of the arguments of
the trigonometric functions in Eqs.\,(\ref{5})\,-\,(\ref{8}), so that all the
 expressions (\ref{14})\,-\,(\ref{19}) can be used in the following in the 
parametrized form. Thus, 
the resonance condition (\ref{14}) can be rewritten as
%------------------------------------------------------25------------------------------
\begin{equation}
{\tan(kr) \over k}= {\varepsilon^{\mu/2} \over \sqrt{-a_1} }\cot\!\left( \sqrt{-a_1}\, 
\varepsilon^{1-\mu/2}\right) + {\varepsilon^{\nu/2} \over \sqrt{-a_2} }\cot\!\left(
\eta\sqrt{-a_2}\, \varepsilon^{1-\mu +\nu/2}\right),
\label{25}
\end{equation}
%------------------------------------------------------25-----------------------------
well defined on the sets $\Omega$ and $B_j$, $j=0,1,2$. 
This representation defines the equation with respect to the intensities $a_1$ and $a_2$
that depends on  the rate of the
distance shrinking either to $r=0$ (single-point interactions) or to $r= n\pi$, $n \in \N$
(double-point interactions). In the present work, we restrict ourselves to the case of 
single-point interactions and therefore assume either $r \equiv 0$ or $r \to 0$. 

%=======================================================4.2============================
\subsection{Resonance conditions and their splitting}
%=======================================================4.2================================

For the first type of point interactions we assume 
 that  $r \to 0$ in such a way that for any $c >0$,
%----------------------------------------------------26------------------------------------
\begin{equation}
{\sin(kr) \over k}\varepsilon^{1-\mu} \to c.
\label{26}
\end{equation}
%------------------------------------------------------26-----------------------------------
Then on the $\Omega$-set and its limiting sets $B_0, B_1, B_2$, the resonance condition (\ref{14}) 
rewritten in the asymptotic form (\ref{25})  reduces to
%------------------------------------------------------27-------------
\begin{eqnarray}
c= \left\{ \begin{array}{llll}
 -\left( 1 / a_1 + 1 / \eta a_2 \right) &   \mbox{for}~ \Omega ,  \\
 \cot\!\sqrt{-a_1} / \sqrt{-a_1} +  \cot(\eta \sqrt{-a_2}\,) / \sqrt{-a_2} 
   & \mbox{for}~ B_0 ,\\
 \cot(\eta \sqrt{-a_2}\,) / \sqrt{-a_2}\, - 1  / a_1 & \mbox{for}~ B_1 ,\\
\cot\!\sqrt{-a_1} / \sqrt{-a_1}\, - 1 / \eta a_2 & \mbox{for}~ B_2.
\end{array} \right. 
\label{27}
\end{eqnarray} 
%-------------------------------------------------------27------------
Similarly, for the second type of point interactions we 
 assume that  $r \to 0$ in such a way that for any $c_0 >0$,
%----------------------------------------------------------28---------------------------------
\begin{equation}
{\sin(kr) \over k} \varepsilon^{2(1-\mu)} \to c_0 .
\label{28}
\end{equation}
%-----------------------------------------------------------28--------------------
The comparison of the limits  (\ref{26}) and (\ref{28})
results in $c=0$ in Eqs.\,(\ref{27}) for the second  type of interactions. 
For the third type the $r \to 0 $ limit is performed in such a way 
that $c_0 =0$ in (\ref{28}), so that for this type  $c=c_0=0$. 
Note also that the linearization of the right-hand  side in (\ref{27}) 
for $B_0$ with respect to $a_1$ and $a_2$ 
 results in the corresponding right-hand expressions 
 for $B_1$, $B_2$ and $\Omega$.

\begin{figure}
\centerline{\includegraphics[width=1.0\textwidth]{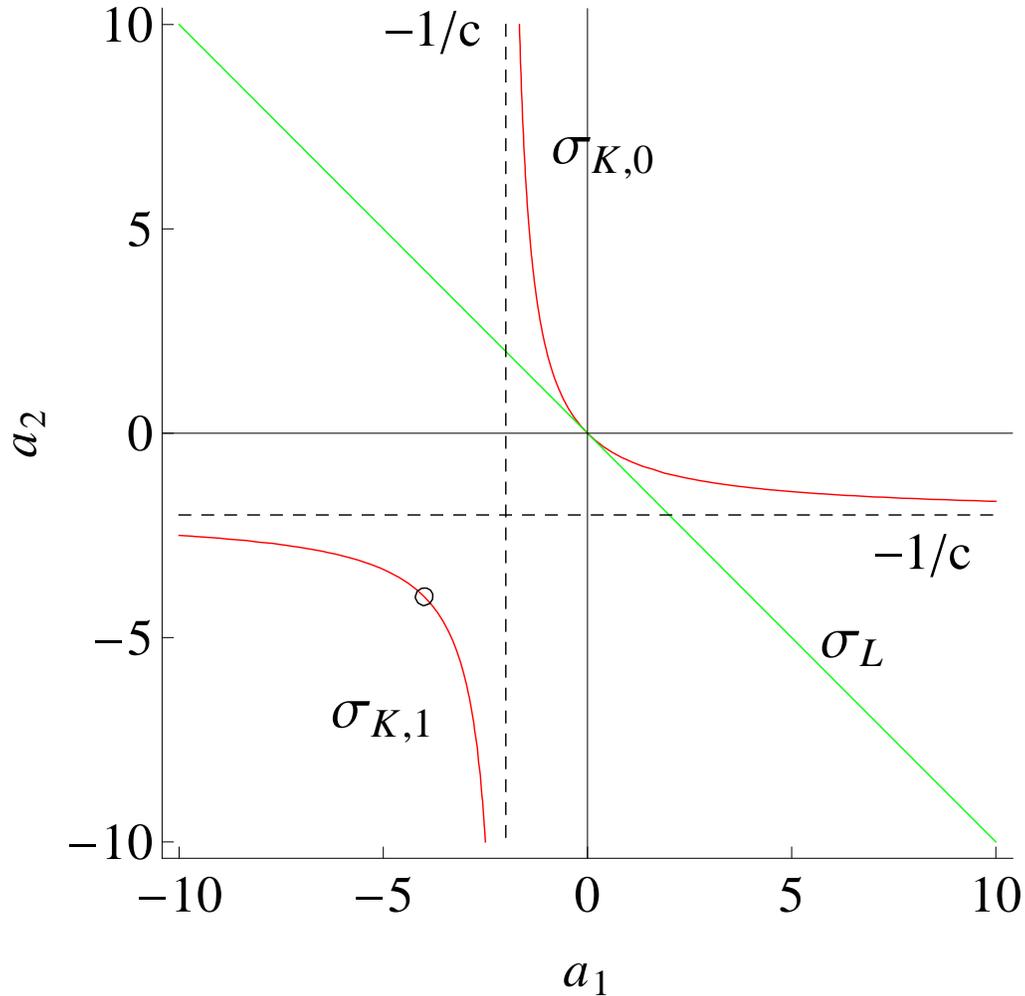}}
%\vspace{2pt}
\caption{
Two disconnected  curves ($\sigma_{K,0}$ and $\sigma_{K,1}$, red lines) 
as a solution of Eq.\,(\ref{27}) for the $\Omega$-set with $c=1/2$ and $\eta=1$ 
forming the resonance set $\Sigma_K$. The curve  $\sigma_{K,0}$
corresponds to the two barrier-well configurations with $h_1h_2 < 0$
($a_1 a_2 <0$) of the potential (\ref{3}), 
while the curve $\sigma_{K,1}$ describes the  
resonance related to the double-well structure.
The point  lying on the line $\sigma_{K,1}$ with 
the coordinates $a_1=a_2= -\, 2/c$ (shown with the empty ball) corresponds to the 
symmetric double-well system with perfect transmission ($\Lambda =-\, I$). 
They belong to the family with the conditions (\ref{12}). The line 
$\sigma_L$ ($a_1 + \eta a_2=0$, green) intersects 
the zeroth resonance curve $\sigma_{K,0}$ only at the origin 
$a_1=a_2 =0$. The coordinates of the asymptotic (dashed) lines are $a_1 = - 1/c$
and $a_2 =- 1/\eta c$. 
When $c \to 0$, the  curve $\sigma_{K,0}$ remains  pinned 
to the origin $a_1 =a_2 =0$  straightening to the line $\sigma_L$ ($\sigma_{K.0}
\to \sigma_L$ as $c \to 0$), while 
the second resonance curve $\sigma_{K,1}$ vanishes escaping to infinity.
}
\label{fig2}
\end{figure}

Thus, the analysis of the resonance sets for all the three types is based 
on Eqs.\,(\ref{27}) with $c \ge 0$. The resonance sets for the $\Omega$-set
as solutions to the first equation (\ref{27})
 are illustrated by Fig.~\ref{fig2} for both $c>0$ (two red curves) and $c=0$ (green line).
%---------------------------------------------fig2-------
%-------------------------------------------------fig2------------- 
The solution with $c>0$ plotted by the two (red) curves 
 $\sigma_{K,0} $ and $ \sigma_{K,1}$ forms  the resonance set $\Sigma_K
 = \sigma_{K,0} \cup \sigma_{K,1}$.
The  resonance curve $\sigma_{K,0}$ appears to be  ``pinned'' to the origin
$a_1=a_2=0$. It can be  considered as  a background 
 branch of the  resonance set and therefore we call it the {\it zeroth} resonance
curve. In the limit as $c \to 0$, 
the curve $\sigma_{K,1}$  vanishes ``escaping'' to infinity, while 
 the zeroth branch straightens to the line (green in Fig.\,\ref{2})
%----------------------------------------------------29---------------------------
\begin{equation}
\sigma_{\!L} := \{ (a_1, a_2) \in \R^2 ~\vert~a_1 +\eta a_2 =0\},
\label{29}
\end{equation}%-----------------------------------------29---------------------------
called in the following the resonance set for the second and the third types of interactions. 

\begin{figure}
\centerline{\includegraphics[width=1.0\textwidth]{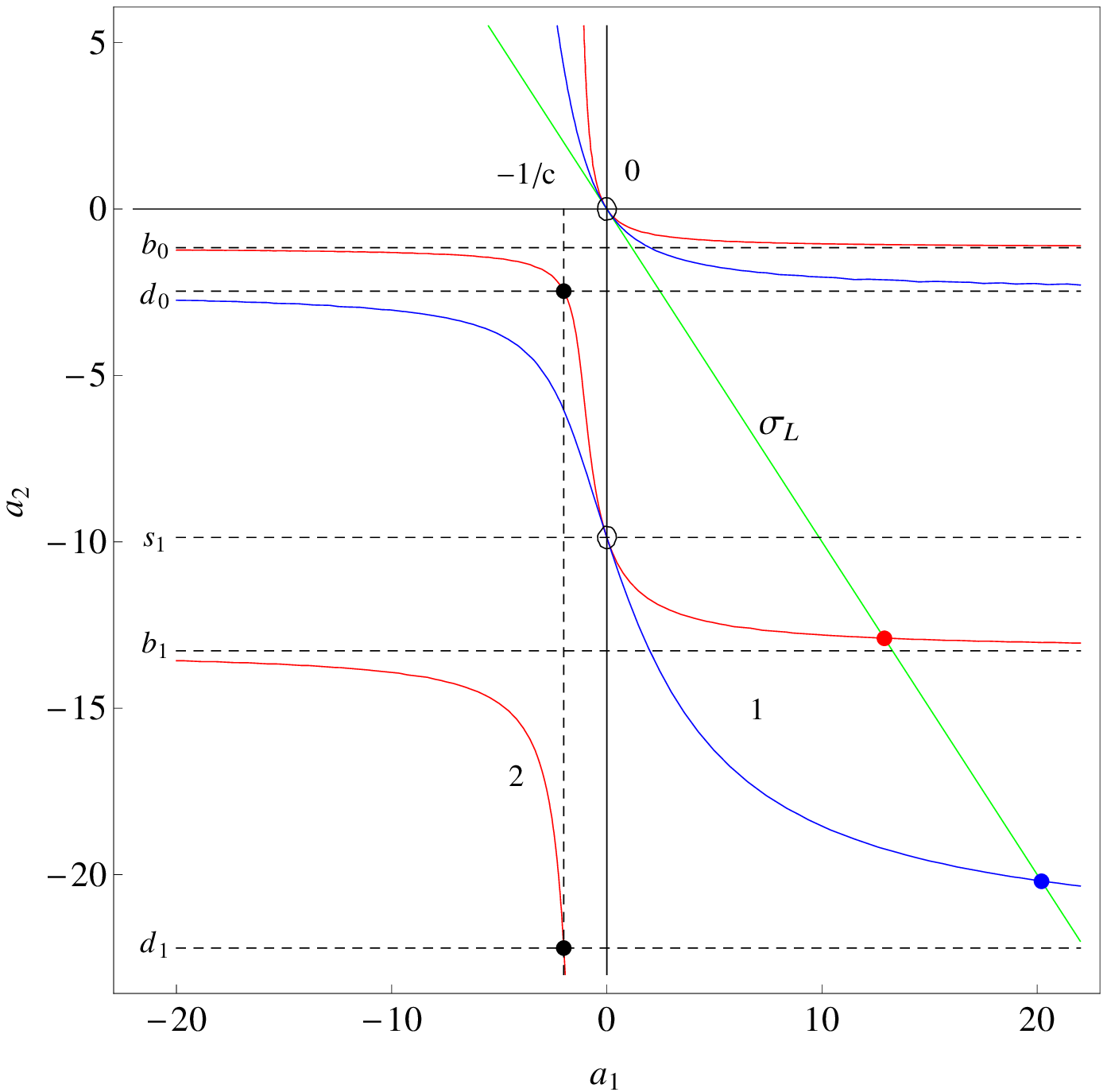}}
%\vspace{2pt}
\caption{ The first three (marked with $n=0,1,2$ 
resonance curves $\sigma_{c>0,n}(B_1)$  (red) and $\sigma_{c=0,n}(B_1)$  (blue)
as solutions to Eq.\,(\ref{27}) for $B_1$ plotted at $\eta =1$ 
with  $c=1/2$  and $c=0$, respectively.
The  curve $\sigma_{c>0,2}(B_1)$ (red) is  depicted partially.
The points $(-1/c, d_{n-1})$ with $n = 1,2$ are shown with the filled (black) balls.
The points shown as the intersection of the first detached resonance curves 
$\sigma_{c>0,1}(B_1)$ (red) and $\sigma_{c=0,1}(B_1)$ (blue) with 
the $\sigma_{\!L}$-line ($a_1 + \eta a_2 =0$, green) belong to 
the resonance sets $\Sigma_{\gamma \delta'}(B_1) $ for 
 the potential $\gamma \delta'(x)$, where $\gamma = \eta a_1/2$
[see Eq.\,(\ref{23}) for $B_1$]  and $a_1$'s are solutions to the equation 
$\sqrt{\eta a_1} \cot\!\sqrt{\eta a_1} = 1 +c a_1$ [see Eq.\,(\ref{27}) for $B_1$]
with $c=1/2$ (red) and $c=0$ (blue).
 }
\label{fig3}
\end{figure}

Next, as follows from the  set of Eqs.\,(\ref{27}) for both the cases with $c>0$ and 
$c=0$, while approaching the limiting sets
$B_j$, $j=0,1,2$, within the open set $\Omega$, the {\em splitting} or 
{\em furcation} of the resonance 
sets $\Sigma_K$ ($c >0$) and $\sigma_L$ ($c=0$) happens and this effect is clearly illustrated by 
Figs.\,\ref{fig3}\,-\,\ref{fig5}.  As shown in these figures, 
 each of Eqs.\,(\ref{27}) for $B_j$, $j=0,1,2$,
admits a countable set of solutions in the form of curves on the 
$\{a_1, a_2 \}$-plane. 
 %------------------------fig3----------------------------
%-------------------------fig3------------------------------------ 
 %------------------------fig4----------------------------
\begin{figure}
\centerline{\includegraphics[width=1.0\textwidth]{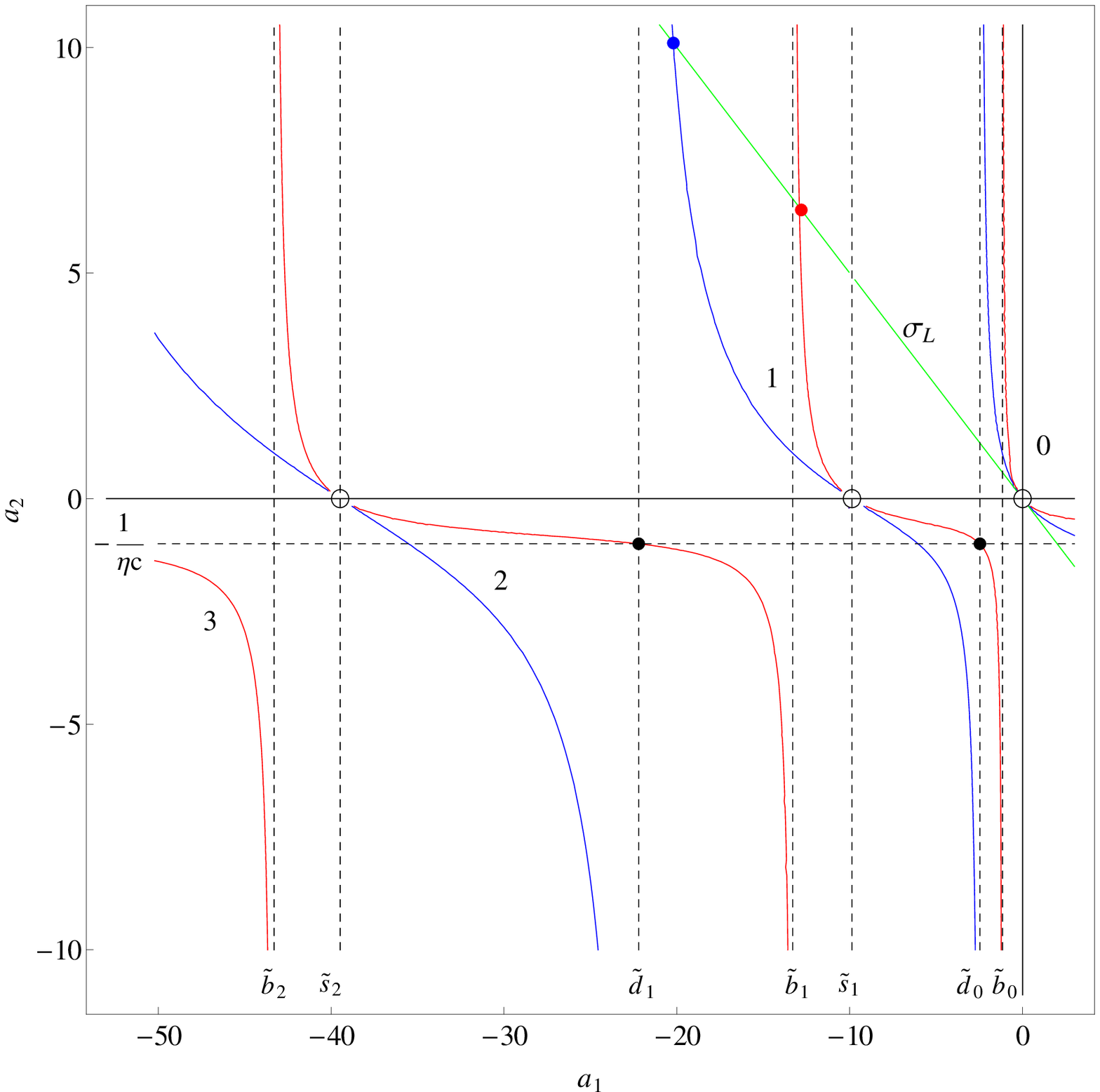}}
%\vspace{2pt}
\caption{ The first four (marked with $n=0,1,2, 3$ 
resonance curves $\sigma_{c>0,n}(B_1)$  (red) and $\sigma_{c=0,n}(B_1)$  (blue)
as solutions to Eq.\,(\ref{27}) for $B_2$ plotted at $\eta=2$ with $c=1/2$  and $c=0$, respectively.
The  curve $\sigma_{c>0,3}(B_2)$ (red) is  depicted partially.
The points $(d_{n-1}, -1/\eta c)$ with $n = 1,2$ are shown with the filled (black) balls.
The points shown as the intersection of the first detached resonance curves 
$\sigma_{c>0,1}(B_2)$ (red) and $\sigma_{c=0,1}(B_2)$ (blue) with 
the $\sigma_{\!L}$-line ($a_1 + \eta a_2 =0$, green) belong to 
the resonance sets $\Sigma_{\gamma \delta'}(B_2) $ for 
 the potential $\gamma \delta'(x)$, where $\gamma =  a_1/2$
[see Eq.\,(\ref{23}) for $B_2$]  and $a_1$'s are solutions to the equation 
$\sqrt{-\, a_1} \cot\!\sqrt{-\,  a_1} = 1 - c a_1$ [see Eq.\,(\ref{27}) for $B_2$]
with $c=1/2$ (red) and $c=0$ (blue).
 }
\label{fig4}
\end{figure}
%-------------------------fig4------------------------------------ 
 %------------------------fig5----------------------------
\begin{figure}
\centerline{\includegraphics[width=1.0\textwidth]{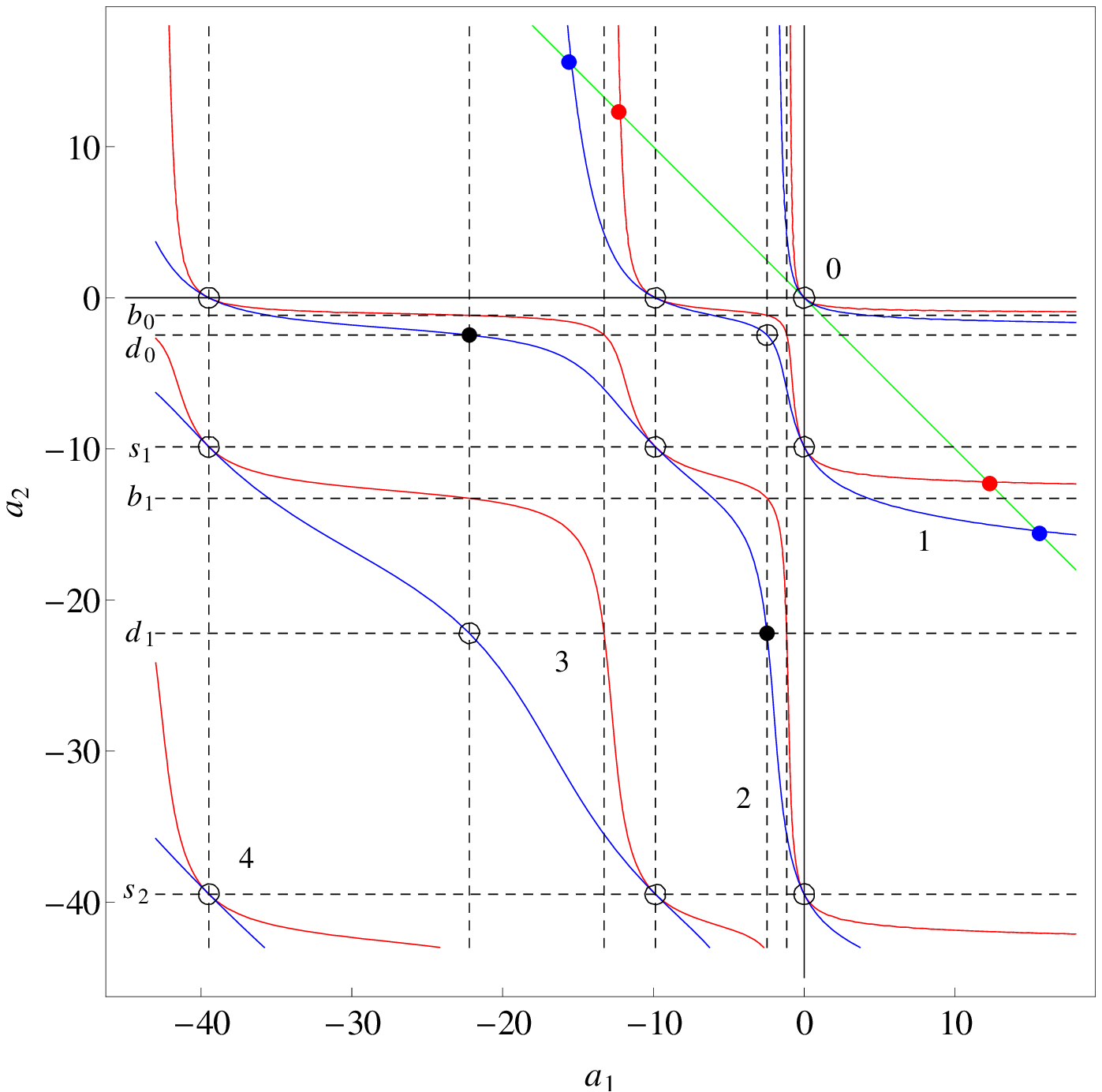}}
%\vspace{0.1pt}
\caption{ 
The first five (marked with $n=0,1,2,3,4$) pairs of the 
resonance curves $\sigma_{c>0,n}(B_0)$  (red) and $\sigma_{c=0,n}(B_0)$ (blue)
as solutions to Eqs.\,(\ref{27}) for $B_0$ at $\eta =1$
 with  $c=1/2$  and $c=0$, respectively. The values $a_1 = \tilde{b}_n,\, \tilde{d}_n, 
\, \tilde{s}_n$  (not shown) correspond to $a_2 = b_n,\, d_n, \, s_n$ placed vertically. 
 The  characteristic points ($\tilde{b}_0,d_0$) and ($\tilde{d}_0,b_0$)
for $n=1$, ($\tilde{b}_0,d_1$), ($\tilde{d}_0,b_1$), ($\tilde{b}_1,d_0$) 
and ($\tilde{d}_1,b_0$) for $n=2$, ($\tilde{b}_1,d_1$)
and ($\tilde{d}_1,b_1$) for $n=3$ that belong to the set $\Sigma_{c>0}(B_0)$
can be seen as the intersection of the corresponding 
vertical and horizontal lines depicted with the dashed lines. The two points 
($\tilde{d}_0, d_1$) and ($\tilde{d}_1, d_0$) shown with the 
black filled balls and lying on the  curve  $\sigma_{c=0,2}(B_0)$ are the 
limits of the pairs  ($\tilde{b}_0,d_1$), ($\tilde{d}_0,b_1$) and ($\tilde{b}_1,d_0$),
 ($\tilde{d}_1,b_0$) as $c \to 0$, respectively. 
   }
\label{fig5}
\end{figure}
%-------------------------fig5------------------------------------ 
These resonance  curves can be numbered by $n =0, 1, \ldots $ and 
we denote them  as $\sigma_{c,n}(B_j)$ for $c \ge 0$, 
which depend on the boundary sets $B_j$, $j=0,1,2$. 
Hence the total resonance sets become as the set functions:
%---------------------------------------30-------------------------------------------
\begin{equation} 
\Sigma_{c}(B_j): =  \cup_{n=0}^\infty \, \sigma_{c, n}(B_j),~~~j=0,1,2.
 \label{30}
\end{equation}
%----------------------------------------30----------------------------------
 The set of the curves with $n =1,2, \ldots  $ may be considered as
the {\it detachment}  from the  zeroth curves $\sigma_{c, 0}(B_j)$.
The comparison of Figs.\,\ref{fig3}\,-\,\ref{fig5} with Fig.\,\ref{fig2}
clearly illustrates the splitting of the resonance sets $\Sigma_{K}$
and $\sigma_{\!L}$ into $\Sigma_{c>0}(B_j)$ and $\Sigma_{c=0}(B_j)$,  respectively. 
Here the zeroth curves $\sigma_{c>0,0}(B_j)$ are deformed a bit if compared 
with the $\sigma_{\!K,0}$-curve shown by the red line in Fig.\,\ref{fig2}.
The location of the split resonance sets on the $\{a_1, a_2 \}$-plane 
for each set 
$B_j $ is described below. The characteristic points on this plane are given in terms of
the $(n +1)$th root (denoted by $b_n = b_n(\eta)$, $b_n\vert_{\eta =1} =: \tilde{b}_n$, 
$n =0, 1, \ldots$) of the equation 
%-----------------------------------------31----------------------------
\begin{equation}
\cot\!\left(\eta \sqrt{-\, b}\,\right)=c\sqrt{-\, b}\,, ~~~-\, \infty < b <0,
\label{31}
\end{equation}
%-----------------------------------------31-------------------------------
and the points 
%---------------------------------------------------32----------------------------------
\begin{equation}
\left. \begin{array}{ll}
d_n=d_n(\eta) := - \, [(n+1/2)\pi/\eta]^2, ~~&  d_n\vert_{\eta =1} =: \tilde{d}_n \\ 
s_n=s_n(\eta)  := -\, (n\pi/\eta)^2, ~~& s_n\vert_{\eta =1} =: \tilde{s}_n 
\end{array} \right.
\label{32}
\end{equation}
%--------------------------------------------------32-----------------------------------
being the solutions of the equations $\cos\!\left(\eta \sqrt{-a_j}\,\right)=0$ 
and $\sin\!\left(\eta\sqrt{-a_j}\, \right)=0$ 
($j=1,2$), respectively. Note that the root ${b}_n$, $n=1,2, \ldots$, 
is  found in the interval $-\, [(n+1/2)\pi/\eta]^2 < {b}_n(\eta) < -\, (n\pi/\eta)^2$,
where $b_n \to d_n$ as $c \to 0$.
The intersection of the $\sigma_{\!L}$-line with the resonance sets $\Sigma_c(B_j)$
defines the discrete point set for the $\gamma \delta'$-potential, where 
the coupling constant $\gamma$ is given by Eqs.\,(\ref{23}) with $a_1 = -\eta a_2$ and $c \ge 0$
satisfying Eqs.\,(\ref{27}). In Figs.\,\ref{fig3}\,-\,\ref{fig5},  the 
red curves belong to $c >0$ and the blue ones to $c=0$. 

{\em Description of the resonance sets $\Sigma_c(B_1)$ 
plotted in Fig.\,\ref{fig3}}:
 For the $B_1$-set, the zeroth resonance curve is located in
the region $\{ -1/c < a_1 \le 0,\, 0 \le a_2 < \infty \} \cup
\{ 0 \le a_1 \le \infty,\, {b}_0 < a_2 \le 0 \}$. 
The asymptotics of the curves with $n =1,2, \ldots  $ 
 are ($a_1 \to -\, \infty,\, a_2 \to {b}_{n-1}$)
and ($a_1 \to +\, \infty,\, a_2 \to {b}_{n}$). 
  Each of these curves  passes through the points 
$ \left( -1/c, d_{n-1} \right)$ and $\left( 0, s_n \right)$. 

{\em Description of the resonance sets $\Sigma_c(B_2)$ 
plotted in Fig.\,\ref{fig4}}:
 For the $B_2$-set, the zeroth resonance curve is located in
the region $\{ \tilde{b}_0  < a_1 \le 0,\, 0 \le a_2 <  \infty \} \cup
\{  0 \le  a_1 <  \infty , \, -1/\eta c < a_2 \le 0 \}$.
The asymptotics of the  curves with $n =1,2, \ldots  $ 
 are ($a_1 \to \tilde{b}_{n}, \, a_2 \to +\, \infty $)
and ($ a_1 \to \tilde{b}_{n-1}, \,  a_2 \to +\, \infty $). 
  Each of these  curves  passes through the points 
$ \left(\tilde{d}_{n-1}, \, -1/\eta c  \right)$ 
and $\left( \tilde{s}_n,\, 0 \right)$. 

{\em Description of the resonance sets $\Sigma_c(B_0)$ 
plotted in Fig.\,\ref{fig5}}:
For the $B_0$-set, the zeroth resonance curve is located in
the region $\{\tilde{b}_0  < a_1 \le 0,\, 0 \le a_2 <  \infty \} \cup
\{  0 \le  a_1 <  \infty , \, b_0  < a_2 \le 0 \}$.
The asymptotics of the $\sigma_{c,n}(B_0)$-curves
are $(a_1 \to \tilde{b}_n,\, a_2 \to + \infty)$ 
and $(a_1 \to + \infty,\, a_2 \to b_n)$, 
$n=0,1, \ldots$.  Each detached curve 
passes through the characteristic points $(\tilde{s}_i,\, s_j)$ with $i+j =n$ and 
$(\tilde{b}_i,\, d_j)$, $(\tilde{d}_i,\, b_j)$ with $i+j =n-1$. 
 
%==============================================4.3=====================================
\subsection{Splitting of the first type of  interactions }
%=================================================4.3=================================

Using the parametrization (\ref{20}), from 
the asymptotic representation of the diagonal elements of the 
$\bar{\Lambda}$-matrix given by the limits (\ref{15}), we obtain
 the following  expressions for $\theta : =\lambda_{11}= \lambda_{22}^{-1} $:
%---------------------------------------------------33---------------
\begin{equation}
 \theta = \left\{ \begin{array}{llll}
1 +ca_1 = \left( 1+ \eta c a_2 \right)^{-1} = -\, a_1 / \eta a_2  &   \mbox{for}~ \Omega , \\
\left( \cos\!\sqrt{-a_1} - c \sqrt{- a_1 }\, \sin\!\sqrt{-a_1} \right)/ 
\cos\left(\eta \sqrt{-a_2}\, \right) & \\
=   \cos\!\sqrt{-a_1}\, \left[\, \cos\left( \eta \sqrt{-a_2}\, \right) 
- c \sqrt{- a_2}\, \sin\left( \eta \sqrt{-a_2}\, \right)  \right]^{-1} &\\
~~~~~~~~  = - \sqrt{a_1 / a_2}\, \sin\!\sqrt{-a_1} /  \sin(\eta \sqrt{-a_2} \,)
   & \mbox{for}~ B_0 ,\\
( 1 +ca_1) / \cos\left(\eta \sqrt{-a_2}\,\right)  &\\
~~~~~~~~ = \left[ \cos\left( \eta \sqrt{-a_2}\, \right) 
- c \sqrt{- a_2}\, \cos\left(\eta \sqrt{-a_2}\, \right) \right]^{-1} &\\
  ~~~~~~~~ =  a_1 /   \sqrt{-a_2}\,  \sin(\eta \sqrt{-a_2}\,)   & \mbox{for}~ B_1 ,\\
 \cos\!\sqrt{-a_1} - c \sqrt{- a_1 }\, \sin\!\sqrt{-a_1}=  \cos\!\sqrt{-a_1}/
( 1+\eta c a_2)   & \\ 
~~~~~~~~ =  \sqrt{-a_1}\, \sin\!\sqrt{-a_1} / \eta  a_2  & \mbox{for}~ B_2
\end{array} \right. 
\label{33}
\end{equation} 
%-----------------------------------------------------33--------------
in the limit as $\varepsilon \to 0$ for the first type of interactions ($c>0$).
This single-point interaction realized on the $\Omega$-set under the assumption (\ref{26}) is referred in the 
following to as  the  resonant-tunneling  $\delta'_K$-interaction.
Setting here
%------------------------------------------------------34-----------------------------
\begin{equation}
 \theta = -\, { a_1 \over \eta a_2}  = { 2+\gamma \over 2-\gamma }\,,
\label{34}
\end{equation}
%-------------------------------------------------------34---------------------------------
we obtain   the limit transmission matrix $\Lambda$ in the  form of (\ref{2}).
 Under the assumption (\ref{34}), 
we obtain Kurasov's $\delta'$-interaction with the intensity 
$\gamma \in \R \setminus 
\{ \pm 2\}$ defined in the distributional sense 
on the space of discontinuous at $x=0$ test functions. For this case one can find the resonance values of $a_1$ and $a_2$  as functions of the strength $\gamma$:
%---------------------------------------------35-----------------------------------
\begin{equation}
a_1 =  {2\gamma \over c_1 ( 2- \gamma) }~~~\mbox{and}~~~
a_2 = -\,  {2\gamma \over \eta c_1 (2+  \gamma)}\,.
\label{35}
\end{equation}
%----------------------------------------------35-----------------------------------
The barrier-well structure corresponds to the interval
$-2 < \gamma < 2$ ($a_1 >0,~a_2<0$ for $-2 < \gamma <0$ and $a_1 <0,~a_2>0$ 
for $0 < \gamma < 2$), whereas beyond this interval ($2< |\gamma|< \infty$), we have the double-well configuration. 
 The boundary conditions beyond the resonance set $\Sigma_K$ are of the Dirichlet type: $\psi(\pm 0)=0$.

Thus the  effect of splitting the $\delta'_K$-interaction occurs while approaching the
limiting $B_j$-sets from the $\Omega$-set. 
On these sets, the limit transmission matrix  $\Lambda$ is of the form (\ref{2})
where the element $\theta$ is determined by Eqs.\,(\ref{33}) defined on the resonance sets $\Sigma_{c>0}(B_j)$, $j=0,1,2$. 
We denote these point interactions as $\delta'_{c>0}(B_j)$ in despite of 
they are no more the point dipoles (double-well configurations are 
present together with barrier-well ones). Schematically, we denote this 
type of splitting as the mapping $\delta'_K \to \delta'_{c>0}(B_j),~j=0,1,2$. 
Similarly, outside the resonance sets  $\Sigma_{c>0}(B_j)$, the limit point interactions satisfy the Dirichlet boundary conditions $\psi(\pm 0)=0$.

%====================================================4.4=============================  
\subsection{Splitting of the second and the third types of interactions }
%========================================================4.4===========================

Similarly, using the parametrization (\ref{20}) in Eqs.\,(\ref{18}) and (\ref{19}), 
we get the  representation of the diagonal elements of the limit 
$\Lambda$-matrix for the second type of interactions 
($\theta := \lambda_{11}= \lambda_{22}^{-1}$):
%------------------------------------------------------36-------------
\begin{equation}
\theta = \left\{ \begin{array}{llll}
 1  &   \mbox{for}~ \Omega , \\
\cos\!\sqrt{- a_1 }/ \cos(\eta \sqrt{-a_2}\,) & \\
~~~~~~~~~~~~ = -\, \sqrt{a_1/a_2}\,
 \sin\!\sqrt{- a_1 }/  \sin(\eta \sqrt{-a_2}\, )  & \mbox{for}~ B_0 ,\\
1 /   \cos(\eta \sqrt{-a_2}\,)= a_1/ \sqrt{-a_2} \,
\sin(\eta \sqrt{-a_2}\, )  & \mbox{for}~ B_1 ,\\
 \cos\!\sqrt{-a_1} = \sqrt{-a_1}\, \sin\!\sqrt{-a_1}/ \eta a_2 & \mbox{for}~ B_2
\end{array} \right. 
\label{36}
\end{equation} 
%------------------------------------------------------36-------------
and
%----------------------------------------------37---------------------
\begin{equation}
\lambda_{21} = \alpha = c_0 \sqrt{a_1 a_2} \left\{ \begin{array}{llll}
 \eta \sqrt{a_1a_2}  &   \mbox{for}~ \Omega , \\
 \sin\!\sqrt{-a_1} \sin(\eta \sqrt{-a_2}  \,) 
   & \mbox{for}~ B_0 ,\\
 \sqrt{-a_1}\,  \sin(\eta \sqrt{-a_2}\,)  & \mbox{for}~ B_1 ,\\
\eta \sqrt{-a_2}\, \sin\!\sqrt{-a_1}  & \mbox{for}~ B_2,
\end{array} \right. 
\label{37}
\end{equation} 
%----------------------------------------------37---------------------
where $a_1$ and $a_2$ satisfy the resonance conditions (\ref{27}) with $c=0$
being defined on the resonance sets $\sigma_L$ (for $\Omega$) 
and  $\Sigma_{c=0}(B_j),~j=0,1,2$. 

The transmission matrix for the point interaction realized on the $\Omega$-set 
corresponds to the potential $\alpha \delta(x)$. In the following we denote 
this point interaction as the $\delta_S$-interaction, 
which is defined on the line $\sigma_{\!L}$. While approaching the 
limit sets $B_j$, the splitting of the $\delta_{\!S}$-interaction occurs resulting in
the point interactions with the transmission matrix of the form 
$\Lambda = \left(\begin{array}{cc} \theta  ~~ ~~~ 0 ~\\
 \alpha ~~ ~~ \theta^{-1} \end {array} \right)$ where  the elements
 $\theta$ and $\alpha$ are given by Eqs.\,(\ref{36}) and (\ref{37}),
 respectively.
We denote these split  interactions as $(\delta\!-\!\delta')_{c=0}(B_j)$
and thus one can use the mapping notation: 
$\delta_S \to (\delta\!-\!\delta')_{c=0}(B_j)$, $j=0,1,2$. 
  
For the third type of interactions $c_0=0$ and therefore $\alpha =0$. 
According to Eq.\,(\ref{36})
for $\Omega$, the transmission matrix on the $\Omega$-set is identically the unit 
($\Lambda = I$) and therefore the realized interactions on this set are 
reflectionless. In the following we denote them as $I_R$, while the point interactions 
realized on the limit $B_j$-sets can be denoted by $\delta'_{c=0}(B_j)$. Thus, 
 one can use the mapping notation: $I_R \to \delta'_{c=0}(B_j)$, $j=0,1,2$. 
 
 As follows from  Eqs.\,(\ref{33}) and (\ref{36}), for the $n$th curve
 passing through the points $(0, s_n)$ in Fig.\,\ref{fig3},  $(\tilde{s}_n,0)$
 in Fig.\,\ref{fig4},  $(\tilde{s}_i,s_j)$, $i+j=n$,  and 
 $\left(\tilde{d}_{2j +1},d_{2j +1}\right)$, $2j +1 =n$, $i,j =0,1, \ldots $, 
 in Fig.\,\ref{fig5}, we have $\theta = (-1)^n$ resulting in  the perfect 
 transmission for the first and the third types of point interactions. 
 These points, which are indicated in the figures  with the empty balls,
 satisfy the conditions (\ref{12}).

%==========================================================5==============================
\section{Geometric  representation of the splitting effect}
%===========================================================5=============================

\begin{figure}
\centerline{\includegraphics[width=1.0\textwidth]{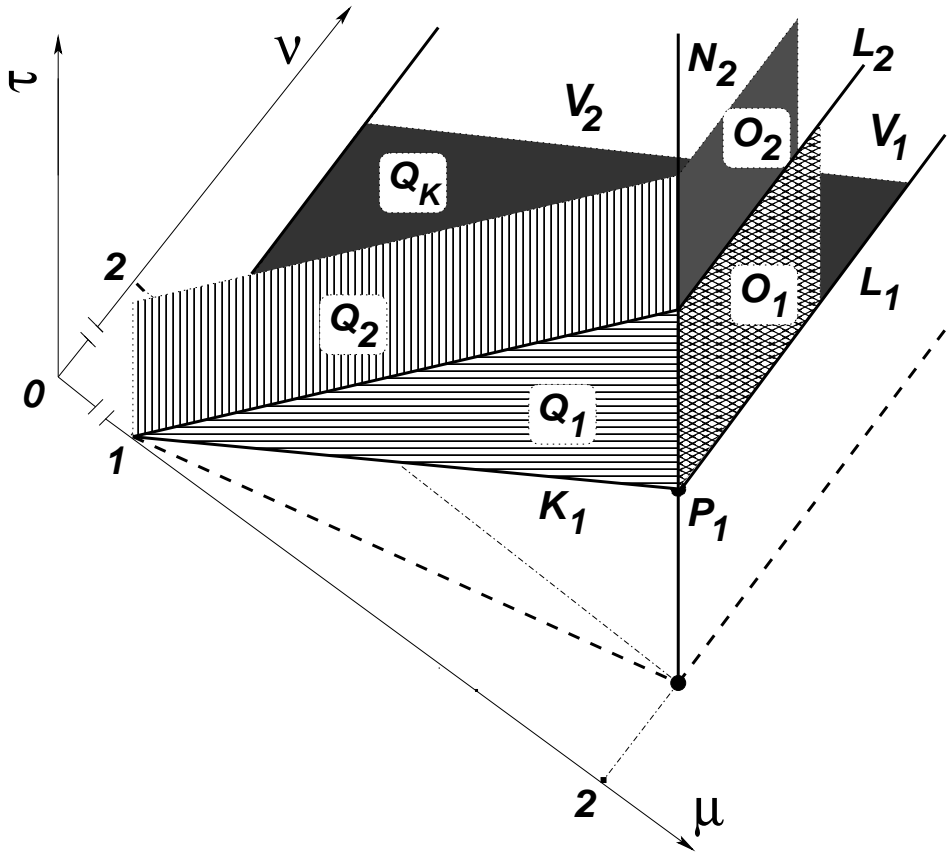}}
%\vspace{0.1pt}
\caption{ Schematics of the trihedral angle surface $S_{\delta'}$ formed by
vertex $P_1$, edges $K_1, L_1, N_1 \cup P_2 \cup N_2$ (line $N_1$ and point $P_2$ shown
in Fig.\,\ref{fig7}),
and planes $Q_1 \cup K_2 \cup Q_2$ (line $K_2 $  shown in Fig.\,\ref{fig7}) 
and $O_1 \cup L_2 \cup O_2$. The interior of the angle is volume set 
$V_1 \cup Q_S \cup V_2$ (plane $Q_S$ shown in Fig.\,\ref{fig7}).
  }
\label{fig6}
\end{figure}

The power-connecting parametrization (\ref{20}) can be extended by adding 
a power parameter for the distance $r$. To this end, 
 we introduce the additional (third) power $\tau $ that describes the rate 
of shrinking the distance $r$ to one point, setting
%--------------------------------------------------38------------------------------------
\begin{equation}
r= c\varepsilon^\tau,~~~c \ge 0,~~\tau >0.
\label{38}
\end{equation}
%--------------------------------------------------38---------------------------------------
Then, adding the third dimension $\tau$ to the $\{\mu, \nu \}$-plane, the dihedral 
angle formed by the sets $B_j \times \{ 0< \tau < \infty \}$, $j=0,1,2$, can be 
considered. The cut off  this angle with the plane $\tau = \mu -1$ forms 
the trihedral angle with the vertex at the point $P_1$ as shown in Fig.\,\ref{fig6}.
In the limit as $\varepsilon \to 0$, the trihedral angle surface 
 appears to be the region where $\delta'(x)$ can be defined 
in the sense of distributions. Therefore we denote this surface by 
$S_{\delta'}$ and the notations of its elements indicated in Figs.\,\ref{fig6} and 
\ref{fig7} will be explained below. 
%---------------------------------------------fig6--------------------------------------
%-------------------------fig6------------------------------------ 

%====================================================5.1====================================
\subsection{Point interactions in the interior of the trihedral angle: \v{S}eba's transition}
%=====================================================5.1===================================

\begin{figure}
\centerline{\includegraphics[width=1.0\textwidth]{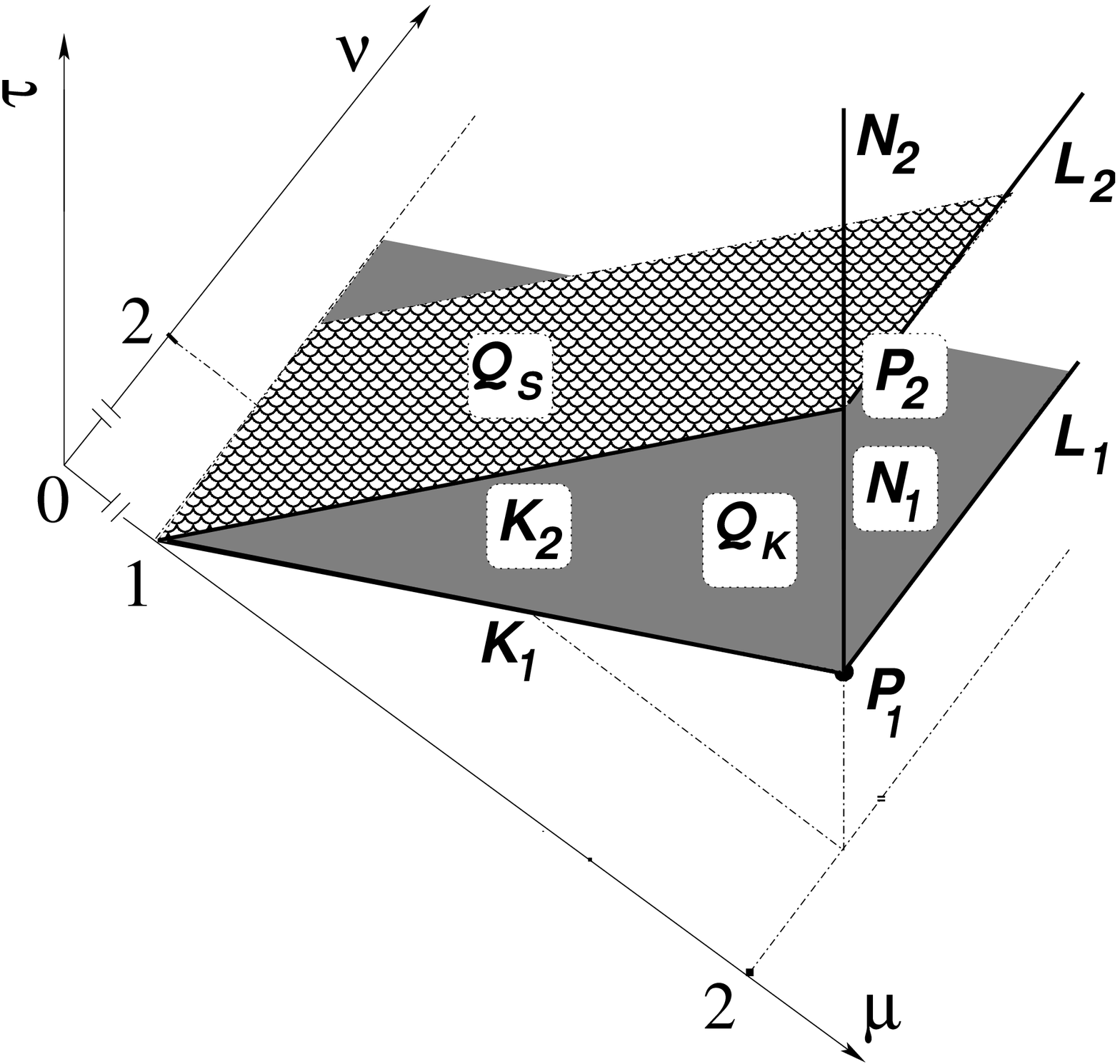}}
%\vspace{0.1pt}
\caption{ Schematics of  plane $Q_K$ with its boundary sets (point $P_1$, lines $K_1$
and $L_1$) and plane $Q_S$ with its boundary sets (point $P_2$, lines $K_2$ and $L_2$).
The edges $K_1$, $L_1$ and  $N_1 \cup P_2 \cup N_2$ form the trihedral angle surface 
$S_{\delta'}$ with vertex at point $P_1$.
  }
\label{fig7}
\end{figure}

Consider first the volume interior of the trihedral angle that consists of the volume regions
$V_1 := \Omega \times \{  \mu -1 < \tau < 2(\mu -1) \}$ and 
$V_2 := \Omega \times \{  2(\mu -1) < \tau < \infty \}$, and the plane 
$Q_S := \Omega \times \{  \tau = 2(\mu -1) \}$ 
 separating these regions (see Fig.\,\ref{fig7}). The point interactions realizing on these sets 
%---------------------------------------------fig7--------------------------------------
%-------------------------fig7------------------------------------ 
appear to be quite 
 different. In the volume set $V_1$, the point interactions
are separated with the boundary conditions of the Dirichlet type $\psi(\pm 0)=0$ and  
therefore the transmission in this region is zero. Contrary, the point interactions in
the volume region $V_2$ are reflectionless (they are denoted by $I_R$) with the resonance
set $\sigma_L$. Note that beyond the $\sigma_L$-set, the point interactions are 
fully non-transparent. Thus, in the interior of the $S_{\delta'}$-surface, the 
$Q_S$-plane with partial transmission serves as a transition region 
from the set $V_1$ of opaque behavior to the volume $V_2$ of perfect transmission. 

In physical terms, the point interactions realized in the volume region $V$ exhibit  
   the transition of transmission that occurs on the  resonance $\sigma_L$-set while
varying the rate of increasing  distance $r$ between the layers in the potential (\ref{3}).
For sufficiently slow squeezing  this distance  [$\mu -1< \tau < 2(\mu -1)$], the limit
point interaction  is opaque, for intermediate shrinking 
[$\tau =2(\mu -1)$] the interaction becomes  partially 
transparent ($\delta$-well) and for fast shrinking [$2(\mu -1) < \tau < \infty $]
the transmission is perfect. In other words, 
the plane $Q_S$ separates the region $V_1$ of full reflection  and the region $V_2$ of
perfect transmission. Therefore the point interaction realized on the plane $Q_S$
may be called the resonant-tunneling   $\delta_S$-interaction
 and that in the region $V_2$ the  resonant-tunneling   reflectionless
$I_R$-interaction. 

Consider now the situation when the thickness of both the layers
in the potential (\ref{3})  squeezes first 
to zero forming the $\delta$-profiles located at $x=0$ and $x=r$, and then 
the $r \to 0$ limit is carried out. As a result, within the parametrizations (\ref{20}) 
and (\ref{38}) at $\eta =1$ 
we get the following asymptotic representation for the potential (\ref{3}):
%---------------------------------------------39--------------------------------
\begin{equation}
V_\varepsilon(x) \to \varepsilon^{1-\mu}
\left[ a_1 \delta(x) +a_2\delta(x-r)\right] = 
(c/r)^\vartheta \left[ a_1 \delta(x) +a_2\delta(x-r)\right]
\label{39}
\end{equation}
%---------------------------------------------39---------------------------------
with $\vartheta : = (\mu -1)/\tau$ in the limit as $r \to 0$. 
This potential has the same form used in 
\cite{s} (see Theorem 3 therein). The transmission matrix of this interaction can 
be computed and, as a result, we find
%-------------------------------------------40----------------------------------------
\begin{equation}
\Lambda_r =  \left( \begin{array}{cc} 1+c^\vartheta r^{1-\vartheta}a_1 ~ & r \\
(c/r)^\vartheta \left(a_1 +a_2 + c^\vartheta r^{1-\vartheta} a_1 a_2\right) ~ & 1 +
c^\vartheta r^{1-\vartheta}a_2  \end{array} \right).
\label{40}
\end{equation}
%-------------------------------------------40--------------------------------------------
It follows from this  matrix that on the line $a_1 + a_2 =0$
at $\vartheta =1/2$ [on the plane $\tau =2(\mu -1)$] we have in the limit as 
$r \to 0$ the resonant 
$\delta$-interaction with the limit transmission matrix 
$\Lambda_{r \to 0} = \left( \begin{array}{cc} 1 & 0 \\
- c a_1^2 ~ & 1 \end{array} \right)$, i.e., the 
result established by \v{S}eba \cite{s}, which agrees with Eqs.\,(\ref{36}) and (\ref{37})
on the $\Omega$-set for $c_0=c$.

In its turn, at $\vartheta =1$ 
(on the plane $\tau =\mu -1$)  the $r \to 0$ limit of the matrix (\ref{40})  reduces to
the limit $\Lambda$-matrix  with the diagonal elements (\ref{33}) for $\Omega$
 corresponding to
Kurasov's $\delta'_K$-interaction. Here the cancellation procedure of divergences 
in the off-diagonal
term results in the resonance condition (\ref{27}) on $\Omega$ with $\eta =1$.
At this condition for all $\vartheta \in (0, 1/2)$ the limit $\Lambda$-matrix is the unit,
while for $\vartheta \in (1/2, 1)$ the limit point interactions are separated
satisfying the  Dirichlet conditions $\psi(\pm 0)=0$.
In physical terms, the value $\vartheta =1/2$ may be called 
a ``transition'' point (at which the
transmission is partial) separating the opaque interaction from
that with perfect transmission. Thus, all the results obtained above for the potential 
(\ref{39}) appear to be in agreement
with those obtained for both the planes  $Q_K$ and $Q_S$:
at $\vartheta =1$ we have the resonance set defined by Eq.\,(\ref{27}) for $\Omega$ 
($\eta =1$) on the plane
$Q_K$, while at $\vartheta=1/2$, i.e., on the plane $Q_S$,  the strength of the
$\delta_S$-interaction is $\alpha = - \, ca_1^2$
 describing the bound state with $\kappa : = \sqrt{-\, E}= - \, \alpha/2 $
($E <0$). Note  that the point interactions with full reflection also
occur on the boundaries of the volume set $V_1$: line 
 $ N_1 := B_0 \times \{1<  \tau < 2 \}$ and planes 
$ Q_1 := B_1 \times \{  \mu -1 < \tau < 2(\mu -1) \} $ and 
$ O_1 := B_2 \times \{1<  \tau < 2 \}$.

%================================================5.2==========================================
\subsection{Splitting of the interactions of the first type}
%====================================================5.2======================================

Using the parametrization (\ref{38}) in Eq.\,(\ref{26}), we find that the point
interactions of the first type are realized on the plane $\tau = \mu -1$. 
More precisely, the $\delta'_K$-interaction is realized on the plane set
$Q_K := \Omega \times \{  \tau = \mu -1 \}$ and its splitting occurs at 
the vertex $P_1 := B_0 \times \{ \tau =1 \}$ and on the edges 
$K_1 := B_1 \times \{  \tau = \mu -1 \}$ and $ L_1 := B_2 \times \{ \tau =1 \}$.
Therefore for the first type one can write the following transitions:
%---------------------------------------------41-----------------------------------
\begin{equation}
Q_K \to {\cal W},~\Sigma_{K} \to \Sigma_{c>0}({\cal W}),~
\delta'_K \to \delta'_{c>0}({\cal W}),~{\cal W}= P_1(B_0), K_1(B_1), L_1(B_2).
\label{41}
\end{equation}
%----------------------------------------------41----------------------------------------

%========================================================5.3=============================
\subsection{Splitting of the interactions of the second type}
%=========================================================5.3============================

Using the parametrization (\ref{38}) in Eq.\,(\ref{28}), we find that the point
interactions of the second type are realized on the plane $\tau = \mu -1$ if $c_0 =c$. 
Here the $\delta_S$-interaction is realized on the plane set
$Q_S $ and its splitting occurs at 
the vertex $P_2 := B_0 \times \{ \tau =2 \}$ and on the edges 
$K_2 := B_1 \times \{  \tau = 2(\mu -1) \}$ and $ L_2 := B_2 \times \{ \tau =2 \}$.
Therefore for the first type one can write the following transitions:
%---------------------------------------------42------------------------------------
\begin{equation}
Q_S \to {\cal W},~ \sigma_{\!L} \to \Sigma_{c=0}({\cal W}),~\delta_S 
\to (\delta-\delta')_{c=0}({\cal W}),~{\cal W}= P_2(B_0), K_2(B_1), L_2(B_2).
\label{42}
\end{equation}
%----------------------------------------------42----------------------------------------

%======================================================5.4================================
\subsection{Splitting of the interactions of the third type}
%=========================================================5.4============================

For the third type of interactions $c_0 =0 $ in Eq.\,(\ref{28}). In this case,
the parametrization (\ref{38}) in Eq.\,(\ref{28}) leads to the existence of point
interactions in the volume set $V_2 $,
which is found above the $Q_S$-plane. The resonance set for these interactions is the same
as for the $\delta_S$-interaction, i.e., $\sigma_L$, but now $\alpha \equiv 0$. 
Hence, due to Eq.\,(\ref{36}) for $\Omega$, this family of resonant tunneling point 
interactions appears to be reflectionless and we denote it by $I_R$. The splitting of
these interactions occurs on the limit sets of $V_2$: edge 
 $N_2 := B_0 \times \{2 <  \tau < \infty \}$ and planes 
$Q_2 := B_1 \times \{  2(\mu -1) < \tau < \infty \}$ and 
$O_2 := B_2 \times \{2 <  \tau < \infty \}.$ The diagonal elements of the limit 
$\Lambda$-matrix are defined by Eqs.\,(\ref{36}) for $B_j$, $j=0,1,2$. Thus, one can
write the mappings:
%---------------------------------------------43------------------------------------
\begin{equation}
V_2 \to {\cal W},~ \sigma_{\!L} \to \Sigma_{c=0}({\cal W}),~I_R \to 
\delta'_{c=0}({\cal W}),~{\cal W}= N_2(B_0), Q_2(B_1), O_2(B_2).
\label{43}
\end{equation}
%----------------------------------------------43---------------------------------------

%==============================================6======================================
\section{Concluding remarks}
%==============================================6========================================

We have studied the pointwise convergence of
the  transmission matrices for the double-layer system  in the squeezing limit
as both the thickness of  the layers  and the
distance between them tend to zero simultaneously. 
Using the $\{\mu, \nu, \tau\}$-parametrization defined by Eqs.\,(\ref{20}) and (\ref{38})
that determines the three-scale  squeezing of the system,   the three types  of  single-point 
interactions with resonant-tunneling behavior have been realized.  
The corresponding resonance sets and the transmission $\Lambda$-matrices have been derived, 
 treating thus the reflection-transmission properties of the double-layer system. 
In particular, on the plane $Q_K$ we have defined 
 Kurasov's $\delta'_K$-interaction \cite{k}
for which the diagonal element $\theta$ in the transmission matrix (\ref{2}) is given by 
Eq.\,(\ref{34}). Under approaching the limiting sets of this plane, the countable splitting
of the $\delta'_K$-interaction occurs that describes the resonant tunneling through the system.
 Unexpectedly, it has been found that \v{S}eba's 
$\delta_S$-interaction introduced in the work \cite{s} can also be included into
the scheme developed in the present paper. 

 For convenience of the presentation, we have used the three-dimensional 
diagram for these powers illustrated by Figs.\,\ref{fig6} and \ref{fig7}, 
where the whole variety of the sets  
corresponds to the family of single-point interactions realized on these sets. 
These sets 
determine how rapidly the squeezing of the distance between the layers 
proceeds in comparison with shrinking the thickness of the layers. 
The  results can be summarized as follows. 
\begin{itemize}
\item
The realization of (both connected and separated) point interactions  occurs in 
 the trihedral angle $V \cup S_{\delta'}$, where $S_{\delta'}$ is the surface on
which the $\delta'$-potential is well defined in the sense of distributions.
\item
The $Q_K $-interaction  realized on the plane $Q_K$
is identified by the transmission matrix of the type (\ref{2}) where the element $\theta$
is given by Eq.\,(\ref{34}). The resonance set $\Sigma_K$ for this interaction, 
being a solution to Eq.\,(\ref{27}) for $\Omega$,
 consists of two curves $\sigma_{K,0}$ and $\sigma_{K,1}$ on the $\{a_1,a_2\}$-plane
as illustrated by Fig.\,\ref{fig2}.  
 \item 
The plane  $Q_S  $  splits  the volume region $V$ into the set $V_1 $
of separated (opaque) interactions satisfying the Dirichlet  conditions 
$\psi(\pm 0)=0$ and the set $V_2 $ 
of the reflectionless interactions denoted by $I_R$.
The $\delta_S$-interaction realized on the  set $Q_S$ 
 is defined by the transmission matrix with the elements (\ref{36}) and (\ref{37})
for $\Omega$. The   resonance sets for both the $\delta_S$- and  $I_R$-interactions are determined  by the line $\sigma_L$.
\item
The splitting phenomenon occurs as the $(\mu, \nu, \tau)$-points on the open 
sets $Q_K,\, Q_S$  and $V_2$ are approaching their limiting sets. These limits
can schematically be presented as the mappings
$$
Q_K \to K_1,\, L_1, \, P_1;~~Q_S \to K_2,\, L_2,\, P_2;~~V_2 \to Q_2,\, O_2,\, N_2.
$$
The zeroth resonance sets $\sigma_{K,0} \subset \Sigma_K$ and $\sigma_L$ as single
curves passing through the origin $a_1 =a_2 =0$
 split into  countable sets.
The splitting of these sets are schematically described as mappings
 by Eqs.\,(\ref{41})\,-\,(\ref{43}).
The comparison of Fig.\,\ref{fig2} with Figs.\,\ref{fig3}\,-\,\ref{fig5}
graphically illustrates the splitting effect. Similarly to the limit 
$\sigma_{K,0} \to \sigma_L$,  the continuous transformation
$ \Sigma_{c>0} \to \Sigma_{c=0}$
takes place as $c \to 0$, despite the sets $K_1,\,L_1$ and $P_1$ are disconnected from
$K_2 \cup Q_2, \, L_2 \cup O_2$ and $P_2 \cup N_2$, respectively.
\item
In the case when the potential (\ref{3}) parametrized by Eqs.\,(\ref{20}) and (\ref{38}) 
converges  to the distribution $\gamma \delta'(x)$ defined on the $S_{\delta'}$-surface, 
the resonance sets $\Sigma_c(B_j)$ are restricted to the countable point sets 
$ \Sigma_{\gamma \delta'}(B_j)= \Sigma_c(B_j) \cap \sigma_L.$

\end{itemize}

The splitting  phenomenon described in the present paper
seems to occur for any multi-layer system. Thus, in the case of $N$ layers
separated equidistantly and determined by intensities $a_1, \ldots, a_N$
(as described in Introduction), the 
$N$-dimensional $S_{\delta'}$-hypersurface for the existence of the distribution
$\delta'(x)$ could be defined. In the $(N+1)$-dimensional open set surrounded by this
surface, the $\delta'_K$-, $\delta_S$- and $I_R$-interactions should be realized
and their countable splitting on some limiting sets located on the 
$S_{\delta'}$-hypersurface seems to take place.  
 Therefore  the approach developed here can 
 be a starting point for further studies on the realization  of point 
interactions in one dimension using  a more general analysis.

\bigskip 
{\bf  Acknowledgments}
\bigskip

The author acknowledges the financial support from the Department of Physics and Astronomy
of the National Academy of Sciences of Ukraine under Project No. 0117U000240. 
 He would like to express gratitude to Yaroslav Zolotaryuk for stimulating
discussions and valuable suggestions.

\end{document}